\documentclass[a4paper,11pt]{article}
\usepackage{jheppub} 
\usepackage{lineno}

\def\d{\mbox{\rm d}}
\def\e{\mbox{\rm e}}

\arxivnumber{2510.01280}

\title{\boldmath Velocity effects slightly mitigating the quantumness degradation of an Unruh-DeWitt detector}

\author[a]{P. H. M. Barros,}
\author[b]{Shu-Min Wu,}
\author[c]{C. A. S. Almeida,}
\author[a]{and H. A. S. Costa}
\affiliation[a]{Departamento de F\'{i}sica, Universidade Federal do Piau\'{i} (UFPI),\\Campus Min. Petr\^{o}nio Portella - Ininga, Teresina - PI, 64049-550 - Brazil}
\affiliation[b]{Department of Physics, Liaoning Normal University, Dalian 116029, China}
\affiliation[c]{Departamento de F\'{i}sica, Universidade Federal do Cear\'{a} (UFC),\\Campus do Pici, Fortaleza - CE, 60455-760 - Brazil}

\emailAdd{phmbarros@ufpi.edu.br, smwu@lnnu.edu.cn, carlos@fisica.ufc.br, hascosta@ufpi.edu.br.}

\abstract{In this work, we investigate the velocity effects on information degradation due to the Unruh effect in accelerated quantum systems (with finite interaction time). We consider a detector moving along a spatial trajectory within a two-dimensional plane, this trajectory is composed of uniform acceleration along one direction, combined with a constant four-velocity component in the plane orthogonal to the acceleration. The quantum systems studied were: accelerated single-qubit, quantum interferometric circuit, and which-path distinguishability circuit. Thus, for non-relativistic velocity regime, we obtained analytical expressions such as transition rates, quantum coherence, visibility, distinguishability, and the complementarity relation. On the other hand, for the ultra-relativistic velocity regime, we saw that the Unruh effect is suppressed and therefore the detector does not respond in this case. Our findings revealed that velocity effects imply mitigation of information degradation, this interesting behaviors happen because of the composite effect of both velocity and acceleration. The results obtained show that the addition of the non-relativistic, transverse and constant motion of an accelerated detector can play a protective role in quantumness in systems at high accelerations, although the effects are very small.}

\begin{document}
\maketitle
\flushbottom

\section{\label{sec:level1}Introduction}

Hawking’s discovery \cite{hawking1975particle} that black holes radiate thermally implies that they eventually evaporate. Semiclassical reasoning indicates that the final quantum state is mixed, due to the entanglement between Hawking radiation and the internal degrees of freedom that vanish upon evaporation \cite{brustein2014horizons, brustein2015constraints, nikolic2024semiclassical}. This conclusion is difficult to avoid, yet it conflicts with the expectation that quantum theory preserves pure states. The tension between these perspectives defines the \emph{black hole information loss paradox} \cite{unruh2017information, hawking2005information}.

In recent years, the relativistic quantum information (RQI) has emerged as a novel research area at the intersection of general relativity and quantum information. Its primary aim is to investigate the role of relativistic effects in quantum information processing protocols~\cite{martin2011relativistic, Mann2012}. Even without considering purely quantum gravity theories, RQI covers a broad spectrum of studies with different scopes. These include the utilization of quantum probes to explore the Unruh and Hawking effects~\cite{hawking1975particle, unruh1976, DeWitt1980, wald1994quantum}, as well as the formulation of quantum protocols on communication and computation~\cite{Adlam2015, Landulfo2016, Martinez2015, Martinez2016, Martinez2020, Martinez2021, wu2022genuine, Tjoa2022, wu2023would, Lapponi2023, wu2023does, wu2024genuinely, wu2025can, liu2025lorentz, liu2025harvesting, tang2025observational, liu2025simulation, li2025does, tang2025can}. Unruh’s seminal work~\cite{Unruh1981Experimental} demonstrated that horizons are not exclusive to gravitational systems, but they also appear in a similar way in laboratories. In particular, sonic horizons in fluids provide an experimentally accessible environment, giving rise to thermal emission similar to Hawking radiation~\cite{visser1998acoustic, gheibi2018magnetoacoustic, tettamanti2019quantum, dave2022hawking, chiofalo2024dissipative}.

Since the last century, the interaction between quantum mechanics and general relativity has revealed conceptually interesting phenomena through the birth of quantum field theory \cite{Nambu1961Dynamical, Dyson1949QED, Feynman1949positrons, Schwinger1948QED}. A paradigmatic example is provided by the Fulling-Davies-Unruh effect~\cite{fulling1973, Davies_1975, unruh1976}, according to which a uniformly accelerated observer in the Minkowski vacuum perceives a thermal spectrum of particles, as if immersed in a heat bath, whereas inertial observers detect no such radiation~\cite{Crispino2008}. Originally formulated by Unruh in 1976~\cite{unruh1976}, this prediction highlights the fundamental role of the Rindler horizon in accelerated frames \cite{rindler1966kruskal, rindler2012essential, falcone2023observing}, which is mathematically analogous to the event horizon of black holes~\cite{birrell1984quantum, kothawala2024noninertial}. Considering the principle of equivalence, similar effects have been observed in the context of the study of black holes; see some recent studies in refs.~\cite{araujo2025non, ARAUJOFILHO2025, araujo2025does, araujo2025particle, ARAUJOFILHO2025117174, Heidari:2025iiv, AraujoFilho:2025jcu}. Consequently, the notion of particle number becomes observer-dependent, as distinct quantization schemes may yield inequivalent definitions of particles~\cite{grove1983notes, wald1994quantum, lancaster2014quantum}.

In order to rigorously analyze this phenomenon, as well as other quantum aspects of field theory in inertial reference frames, the Unruh-DeWitt (UDW) detector was introduced~\cite{unruh1976, DeWitt1980}. This construction constitutes a simplified yet powerful model, wherein a two-level quantum system is coupled to a quantum field. Despite its simplicity, the UDW detector faithfully captures the essential features of how distinct classes of observers interact with quantum fields. It thereby serves as a conceptual probe, enabling one to characterize and quantify the particle content perceived as a consequence of acceleration. Within this framework, one may thus operationally define particles as those excitations registered by an idealized particle detector \cite{davies1984}.

A number of strategies have been put forward to probe the detection of particles associated with the Unruh effect. Among these, approaches based on quantum coherence have recently attracted considerable attention~\cite{tian2012unruh, Wang2016, he2018multipartite, Nesterov2020, Zhang2022, huang2022, Harikrishnan2022, xu2023decoherence, Pedro2024robustness, wu2025can}. In this structure, a UDW detector couples to the quantum field, absorbing quanta, thus modifying its eigenstates and leading to coherence degradation. Several recent investigations have identified conditions under which this loss of coherence is amplified. Such amplification has been reported when the vacuum is modeled as a dispersive medium~\cite{barros2024dispersive}, similarly when a gravitational wave background is present~\cite{barros2024detecting}, and when the scalar field is quadratically coupled with the detector~\cite{barros2025information}. In contrast, our recent results provide evidence that the mass of the scalar field can act as a protective mechanism~\cite{pedro2025mitigating}.

Another strategy for probing this phenomenon relies on the use of quantum interferometric circuits~\cite{Costa2020interferometry, GoodingUnruh2020interferometric, Lopes2021interference, Pedro2024robustness, barros2024detecting, pedro2025mitigating}. Within this framework, the particles generated due to acceleration modify the probability amplitudes that cause degradations in the visibility and interference pattern. Furthermore, interferometric configurations can be adapted to examine matter-like features through the analysis of which-path distinguishability~\cite{Jaeger1995, Englert1996, miniatura2007path, kolavr2007path, bera2015duality}. By simultaneously accounting for both wave-like and particle-like properties, one may derive the complementarity relation within this context. Such a formulation provides a natural setting for investigating how Unruh radiation influences wave-particle duality~\cite{Pedro2024robustness, pedro2025mitigating, barros2025information}.

In the last century, stationary worldlines and vacuum excitation of non-inertial detectors were considered~\cite{Letaw1981stationary}, however, a special type of trajectory stands out where a detector moving along a spatial trajectory in a two-dimensional spatial plane has constant independent magnitudes of both the four-dimensional acceleration and a proper time derivative of the four-dimensional acceleration, so that the direction of the acceleration rotates at a constant rate around a great circle~\cite{abdolrahimi2014velocity}. However, it has recently been shown that superposition of accelerated linear motion and a four-velocity component of acceleration can inhibit Fisher quantum information degradation~\cite{Liu2021relativistic}, quantum work degradation~\cite{Hao2023Quantum}, and the decoherence of a quantum battery quadratic environmental coupling~\cite{mukherjee2024enhancement}.

Motivated by these results, this work investigates the types of effects caused by the four-velocity component in some accelerated quantum systems constructed by UDW detectors that interact for a finite time with a massless scalar field. This work is structured as follows: In Sec.~\ref{sec:level2}, we present the UDW theory considering velocity effects, however, we analytically show the detector response for both velocity regimes: non-relativistic and ultra-relativistic. In Sec.~\ref{sec:level3}, we obtained quantum coherence by considering velocity effects for an accelerated single-qubit. In Sec.~\ref{sec:level4}, we show the results on wave-particle duality considering velocity effects, where we first obtain the visibility and which-path distinguishability and then calculate the complementarity relation. In Sec.~\ref{sec:level5}, we clarify the assumptions and limits of our model, and finally, in Sec.~\ref{sec:level6}, we present our conclusions on the present work.

For convenience, natural units are employed throughout this work by setting $c = \hbar = k_{B} = 1$. Furthermore, we adopt the metric signature $\eta^{\mu\nu} = (+,-,-,-)$.

\section{\label{sec:level2}Unruh-DeWitt theory with velocity effects}

\subsection{The theoretical model}

For the purposes of this study, we begin by analyzing an UDW detector \cite{unruh1976, DeWitt1980}, namely, a point-like system characterized by two distinct energy levels: a ground state $\vert g \rangle$ and an excited state $\vert e \rangle$. Within this framework, the detector’s proper time is denoted by $\tau$, while its trajectory is described by the worldline $x^{\mu}(\tau) = (t(\tau),\mathbf{x}(\tau))$ where $\mu$ labels the
coordinates in space-time. We focus on the scenario in which the detector couples linearly to a massless scalar field $\phi[x(\tau)]$ through its monopole moment operator $\mu(\tau)$ \cite{unruh1976, DeWitt1980, birrell1984quantum, frolov2011introduction}. Accordingly, the Hamiltonian governing the linear interaction between the detector and the scalar field takes the following form:
\begin{eqnarray}
  \mathcal{H}_{\mathrm{int}} = \lambda \chi(\tau)\mu(\tau) \otimes \phi[x(\tau)],
  \label{Hint}
\end{eqnarray}
where $\lambda$ denotes the interaction strength, while $\chi(\tau)$ is the switching function that regulates the activation and deactivation of the detector. Owing to the finite duration of the interaction, this function satisfies the following conditions: $\chi(\tau) \approx 1$ for $|\tau| \ll T$, and $\chi(\tau) \approx 0$ for $|\tau| \gg T$, where $T$ characterizes the effective interaction timescale.

Within the first-order interaction regime, the excitation and de-excitation probabilities are respectively given by $\mathcal{P}^{\pm} = \lambda^2\vert\langle g\vert\mu(0)\vert e\rangle\vert^2 \mathcal{F}^{\pm}$. Here, $\mathcal{F}^{\pm}$ denote the response functions, which encode the information associated with the detector’s dynamical behavior. From a mathematical perspective, the functions $\mathcal{F}^{\pm}$ are
\begin{eqnarray} 
\mathcal{F}^{\pm} &=& \int_{-\infty}^{\infty} \d\tau  \int_{-\infty}^{\infty} \d\tau' \chi(\tau)\chi(\tau') \e^{\pm i\Omega(\tau - \tau')} \mathcal{W}^{+}[x(\tau), x(\tau')],
\label{Response}
\end{eqnarray}
with $\Omega$ being the angular transition frequency. For $\Omega > 0$, the process corresponds to the absorption of a quantum, resulting in the excitation of the detector. Conversely, when $\Omega < 0$, the process corresponds to the emission of a quantum, leading to the de-excitation of the detector. Moreover, $\mathcal{W}^{+}[x(\tau),x(\tau')]$ denotes the Wightman function, defined as $\mathcal{W}^{+}[x(\tau), x(\tau')] = \langle 0_{\mathcal{M}}|\phi[x(\tau)]\phi[x(\tau')]|0_{\mathcal{M}}\rangle$. It is worth noting that this expression remains invariant under time translations for both inertial and uniformly accelerated trajectories. In other words, $\mathcal{W}^{+}[x(\tau), x(\tau')] = \mathcal{W}^{+}(\tau - \tau')$, as discussed in refs.~\cite{letaw1981quantized,padmanabhan1982general}.

Recalling the property satisfied by the Wightman function [see refs.~\cite{letaw1981quantized, sriramkumar1996finite}], namely, that $\mathcal{F}^{\pm}$ are
\begin{align}
    f(u)\left[\e^{-i\Omega u}\, \mathcal{W}^{+}(u)\right]=f\left(-\frac{\partial}{\partial\Omega}\right)\left[\e^{-i\Omega u}\mathcal{W}^{+}(u)\right],
\end{align}
where $f(u)$ denotes an arbitrary function that admits a well-defined Taylor series expansion around $u=0$. Consequently, the asymptotic expression for the transition probability associated with any analytic window function is
\begin{align}
        \mathcal{F}^{\pm} = \chi\left(i\frac{\partial}{\partial \Omega}\right)\chi\left(-i\frac{\partial}{\partial \Omega}\right) \mathcal{F}^{\pm}(\infty).
\end{align}
Therefore, the transition probability $\mathcal{F}^{\pm}(\infty)$ for an infinite time detector reduces to
\begin{align}
    \mathcal{F}^{\pm}(\infty)=\int_{-\infty}^{\infty}\,\d\tau\,\int_{-\infty}^{\infty}\,\d\tau'\, \e^{\pm i\Omega(\tau-\tau')} \mathcal{W}^{+}(\tau-\tau').
\end{align}

We proceed by expanding the function $\chi(\tau)$ as a Taylor series about $\tau=0$, under the assumptions $\chi(0)=1$ and $\chi'(0)=0$, which yields
\begin{align}
    \mathcal{F}^{\pm} \approx \mathcal{F}^{\pm}(\infty)-\chi''(0)\frac{\partial^2\mathcal{F}^{\pm}(\infty)}{\partial\Omega^2}.
\end{align}

Therefore, the transition probability per unit time can be defined as
\begin{align}
    \mathcal{R}^{\pm}\approx \mathcal{R}^{\pm}(\infty)-\chi''(0)\frac{\partial^2 \mathcal{R}^{\pm}(\infty)}{\partial \Omega^2}, 
\end{align}
where
\begin{align}
    \mathcal{R}^{\pm}(\infty)=\int_{-\infty}^{\infty}\, \d(\Delta\tau)\,\e^{\pm i\Omega \Delta\tau} \mathcal{W}^{+}(\Delta \tau),
    \label{Int R infty}
\end{align}
with $ \Delta \tau=\tau-\tau'$. It is noteworthy that the transition probability is sensitive to the derivatives of the window function. Consequently, if the detector is switched on and off abruptly, these contributions can give rise to divergent responses. To circumvent such divergences while preserving invariance under time translations for both accelerated and inertial trajectories, we employ a carefully chosen window function profile. Naturally, these considerations lead us to adopt a Gaussian window function for the UDW detector, which offers technical advantages by allowing precise control over the window width through regulation of the integrals. Accordingly, we assume the Gaussian window function~$\chi(\tau)=\text{e}^{-\frac{\tau^2}{2T^2}}$ \cite{sriramkumar1996finite}. By implementing this profile and restricting the interaction to a finite time interval, the resulting expression for the transition probability per unit time is
\begin{eqnarray}
 \mathcal{R}^{\pm} \approx \mathcal{R}^{\pm}(\infty) + \frac{1}{2T^2}\frac{\partial^2 \mathcal{R}^{\pm}(\infty)}{\partial\Omega^2} + \mathcal{O}(T^{-4}),
 \label{R}
\end{eqnarray}
where $\overline{\mathcal{R}}^{-}$ and $\overline{\mathcal{R}}^{+}$ are the rates of the excitation probability and the de-excitation probability, respectively. Besides, $\overline{\mathcal{R}}^{\pm}(\infty)$ is the transition rate for an infinite time detector ($T \to \infty$).

\subsection{Transition probability rates}

Consider a UDW detector, as described in the previous section, moving in Minkowski spacetime along an unbounded spatial trajectory within a two-dimensional spatial plane. The detector follows a path characterized by a constant squared magnitude of the four-acceleration, $a^2 = a_{\mu} a^{\mu}$, where $a^{\mu} = \d^2 x^{\mu}/\d\tau^2$, as well as a constant magnitude of the timelike proper-time derivative of the four-acceleration, $(\d a_{\mu}/\d\tau) (\d a^{\mu}/\d\tau)$. Furthermore, one component of the four-velocity, $\d y/\d\tau = w$, is maintained constant. That is, the detector moves along the following worldline per unit proper time \cite{Letaw1981stationary, abdolrahimi2014velocity}:
\begin{eqnarray}
    x^{\mu}(\tau) = \left( \frac{a}{\alpha^2} \sinh{(\alpha\tau)}, \frac{a}{\alpha^2}\cosh{(\alpha\tau)}, w\tau, 0\right),
    \label{trajectory}
\end{eqnarray}
where
\begin{eqnarray}
    \alpha = \frac{a}{\sqrt{1+w^2}} > 0,
    \label{alpha}
\end{eqnarray}
with $x^{\mu} = (t, x, y, z)$ being the Minkowski coordinates. In this path, as shown by ref.~\cite{abdolrahimi2014velocity}, the Wightman function corresponding to the trajectory given by Eq.~(\ref{trajectory}) is given by
\begin{eqnarray}
    \mathcal{W}^{+}_{w}(\Delta\tau) = - \frac{\alpha^4}{16\pi^2 a^2} \left[ \sinh^2{\left( \frac{\alpha\Delta\tau}{2} - \frac{i\epsilon\alpha^2}{a}\right)} - \frac{w^2\alpha^4}{4a^2}\Delta\tau^2\right]^{-1},
    \label{wightman function}
\end{eqnarray}
where it can be observed that for $w = 0$, we have $\alpha = a$, and Eq.~(\ref{wightman function}) reduces to the positive Wightman function corresponding to a detector moving along a spatially straight trajectory in the $x$-direction, with a constant magnitude of the four-acceleration, given by
\begin{eqnarray}
    \mathcal{W}^{+}_{0}(\Delta\tau) = - \frac{a^2}{16\pi^2} \left[ \sinh^2{\left( \frac{a\Delta\tau}{2} - i\epsilon a\right)} \right]^{-1}.
    \label{wightman function zero}
\end{eqnarray}

It is worth emphasizing that we can analyze Eq.~(\ref{wightman function}) for the non-relativistic velocity regime ($w \ll 1$) and for the ultra-relativistic velocity regime ($w \gg 1$). Using these regimes is essential to reveal distinct physical behaviors and allow specific analytical analyses of the detector response.

\subsubsection{Non-relativistic velocity regime}

Note that in the present work, the non-relativistic velocity regime ($w \ll 1$) does not refer to a classical description of motion, but rather to the relativistic expansion at low spatial velocities compared to the speed of light, so that corrections to the detector response emerge only from terms of order $w^2$. We now compute the transition probability rates with a finite interaction time for the non-relativistic regime. For this, we derive in detail in Appendix~\ref{App:A} the expansion of the Wightman function [Eq.~(\ref{wightman function})] in the principal order for $w \ll 1$, thus according to Eq.~(\ref{Wightman function w<<1}) we have
\begin{eqnarray}
    \mathcal{W}_{w}^{+}(\Delta\tau) &\approx& -\frac{a^2}{16\pi^2} \Bigg\{ \frac{(1-2w^2)}{\sinh^2{\left(  \frac{a\Delta\tau}{2} - i\epsilon a\right)}}  + \Bigg[\left( \frac{\Delta\tau}{4} - i\epsilon a\right)\sinh{\left(  a\Delta\tau - 2i\epsilon a\right)} + \nonumber\\
    &+& \frac{(a\Delta\tau)^2}{4}\Bigg]\frac{w^2}{\sinh^4{\left(  \frac{a\Delta\tau}{2} - i\epsilon a\right)}} \Bigg\} + \mathcal{O}(w^4),
    \label{wightman non-relativistic}
\end{eqnarray}
and to calculate the detector transition rate, we substitute Eq.~(\ref{wightman non-relativistic}) into Eq.~(\ref{Int R infty}), calculate the contour integrals (for more information see ref.~\cite{abdolrahimi2014velocity}), and defining the following dimensionless parameters $\overline{\mathcal{R}}^{\pm} = \mathcal{R}^{\pm}/\Omega$ and $\overline{a} = a/\Omega$, thus we have the excitation probability rate
\begin{eqnarray}
    \overline{\mathcal{R}}^{-}_{w}(\infty) = \frac{1}{2\pi}\frac{1}{\e^{2\pi/\overline{a}}-1} - F(\overline{a})w^2,
    \label{R- infty}
\end{eqnarray}
where we have a function of acceleration $F(\overline{a})$ given by
\begin{eqnarray}
    F(\overline{a}) = \frac{\overline{a}\e^{2\pi/\overline{a}}}{6(\e^{2\pi/\overline{a}}-1)^2} \left[ 2 + \frac{9}{\overline{a}^2} - \frac{2\pi}{\overline{a}} \left( 1 + \frac{1}{\overline{a}^2}\right) \coth{\left(\frac{\pi}{\overline{a}}\right)}\right],
\end{eqnarray}
and the probability rate of de-excitation
\begin{eqnarray}
    \overline{\mathcal{R}}^{+}_{w}(\infty) = \frac{1}{2\pi}\frac{\e^{2\pi/\overline{a}}}{\e^{2\pi/\overline{a}}-1} - \e^{2\pi/\overline{a}}\, F(\overline{a})w^2,
    \label{R+ infty}
\end{eqnarray}
where the relation given by $\overline{\mathcal{R}}^{+}_{w}(\infty) = \e^{2\pi/\overline{a}}\overline{\mathcal{R}}^{-}_{w}(\infty)$ holds.

In this way, to obtain the transition rates during a finite interaction time, we substitute Eqs.~(\ref{R- infty}) and (\ref{R+ infty}) into Eq.~(\ref{R}), calculating the derivatives, and after several algebraic simplifications, we obtain
\begin{eqnarray}
    \overline{\mathcal{R}}^{-}_{w} &\approx& \overline{\mathcal{R}}^{-}_{w}(\infty) \left\{ 1 - \frac{\pi\,\e^{2\pi/\overline{a}} [\mathcal{A}^{-}_{1}(\overline{a},w) + \mathcal{A}^{-}_{2}(\overline{a},w) + \mathcal{A}^{-}_{3}(\overline{a},w) ]}{\overline{a}^2\sigma^2(\e^{2\pi/\overline{a}}-1)^{3}\,\mathcal{B}(\overline{a},w) }\right\},
    \label{R-}
\end{eqnarray}
where, we have
\begin{eqnarray}
\label{A1-}
    \mathcal{A}^{-}_{1}(\overline{a},w) &=& 30\pi^2 \,\overline{a} \,w^2(-1-3\e^{2\pi/\overline{a}} + 3\e^{4\pi/\overline{a}} + 6\e^{2\pi/\overline{a}}) \nonumber\\
    &+& 4\pi^3w^2 (1 + 11\e^{2\pi/\overline{a}} + 11\e^{4\pi/\overline{a}} + \e^{6\pi/\overline{a}}),
\end{eqnarray}
\begin{eqnarray}
\label{A2-}
    \mathcal{A}^{-}_{2}(\overline{a},w) &=& 2\pi\, \overline{a}^2 (\e^{2\pi/\overline{a}} + 1) \Big\{ 3 + (21 + 2\pi^2)w^2 \nonumber\\ &+& \e^{4\pi/\overline{a}} \Big[3 + (21 + 2\pi^2)w^2 + \e^{2\pi/\overline{a}}((20\pi^2-42)w^2-6)\Big] \Big\},
\end{eqnarray}
\begin{eqnarray}
\label{A3-}
    \mathcal{A}^{-}_{3}(\overline{a},w) &=& -\overline{a}^3 (\e^{2\pi/\overline{a}} - 1) \Big\{ 6 + (9 + 8\pi^2)w^2 \nonumber\\ &+& \e^{4\pi/\overline{a}} [6 + (9 + 8\pi^2)w^2] + 2\e^{2\pi/\overline{a}}[(16\pi^2-9)w^2-6] \Big\},
\end{eqnarray}
\begin{eqnarray}
    \mathcal{B}(\overline{a},w) = 2\pi\overline{a}^3 w^2 - 3\overline{a}^2(\e^{2\pi/\overline{a}}-1) + 9\pi \overline{a} w^2 - 2\pi^2 w^2 \e^{2\pi/\overline{a}}(1+\overline{a}^2) \coth{\left( \frac{\pi}{\overline{a}}\right)},
\end{eqnarray}
and similarly for the de-excitation rate reads
\begin{eqnarray}
    \overline{\mathcal{R}}^{+}_{w} &\approx& \overline{\mathcal{R}}^{+}_{w}(\infty) \left\{ 1 - \frac{\pi\,\e^{2\pi/\overline{a}} [\mathcal{A}^{+}_{1}(\overline{a},w) + \mathcal{A}^{+}_{2}(\overline{a},w) + \mathcal{A}^{+}_{3}(\overline{a},w) ]}{\overline{a}^2\sigma^2(\e^{2\pi/\overline{a}}-1)^{3}\,\mathcal{B}(\overline{a},w) }\right\},
    \label{R+}
\end{eqnarray}
where, we have
\begin{eqnarray}
\label{A1+}
    \mathcal{A}^{+}_{1}(\overline{a},w) &=& -12\pi^2 \,\overline{a} \,w^2(-8\e^{2\pi/\overline{a}} + \e^{4\pi/\overline{a}} + 7\e^{6\pi/\overline{a}}) \nonumber\\
    &+& 16\pi^3 w^2 (\e^{2\pi/\overline{a}} + 4\e^{4\pi/\overline{a}} + \e^{6\pi/\overline{a}} + \e^{6\pi/\overline{a}}),
\end{eqnarray}
\begin{eqnarray}
\label{A2+}
    \mathcal{A}^{+}_{2}(\overline{a},w) &=& 2\pi\, \overline{a}^2  \Big\{ 3 + 3\e^{8\pi/\overline{a}}w^2 + \e^{6\pi/\overline{a}}[3 + (33 + 8\pi^2)w^2] \nonumber\\
    &+& \e^{2\pi/\overline{a}} [-3 + (39 + 8\pi^2)w^] + \e^{4\pi/\overline{a}}[(32\pi^2-75)w^2-3)] \Big\},
\end{eqnarray}
\begin{eqnarray}
\label{A3+}
    \mathcal{A}^{+}_{3}(\overline{a},w) &=& -3\overline{a}^3 (\e^{2\pi/\overline{a}} - 1) \Big\{ 2 + 3\e^{6\pi/\overline{a}}w^2 \nonumber\\ &+& \e^{4\pi/\overline{a}}[2+ (-6 + 8\pi^2)w^2] + \e^{2\pi/\overline{a}} [-4 + (3 + 8\pi^2)w^2] \Big\}.
\end{eqnarray}

It is important to note that when we take $w= 0$ in Eqs.~(\ref{R-}) and (\ref{R+}), we obtain the standard case without velocity effects \cite{sriramkumar1996finite, Pedro2024robustness}. The expressions Eq.~(\ref{R-}) and Eq.~(\ref{R+}) present slight modifications of the thermal spectrum, moreover, the dependence of $w$ highlights the velocity effect on the detector transition probability rates.

\subsubsection{Ultra-relativistic velocity regime}

Now, let us examine the case in which the detector follows the trajectory given in Eq.~(\ref{trajectory}), in the ultra-relativistic regime, namely in the limit $w \gg 1$. In this context, we derive in detail in Appendix~\ref{App:A} the expansion for $w \ll 1$ of the Wightman function, as expressed in Eq.~(\ref{wightman function}), which is written as
\begin{eqnarray}
    \mathcal{W}_{w}^{+}(\Delta\tau) &\approx& -\frac{a^2}{16\pi^2}\frac{1}{w^4} \Bigg[ \sinh^2{\left(  \frac{a\Delta\tau}{2w} - \frac{i\epsilon a}{w^2}\right)}\Bigg]^{-1}  + \mathcal{O}(w^{-6}),
    \label{wightman function ultra}
\end{eqnarray}
and it is easy to see that the detector responses obtained through contour integration of Eq.~(\ref{wightman function ultra}) is given by
\begin{eqnarray}
    \overline{\mathcal{R}}^{-}_{w}(\infty) &=& \frac{1}{2\pi}\frac{1}{(\e^{2\pi w/\overline{a}}-1)} \frac{1}{w^2},\label{R- infty ultra}\\
    \overline{\mathcal{R}}^{+}_{w}(\infty) &=& \frac{1}{2\pi}\frac{\e^{2\pi/\overline{a}}}{(\e^{2\pi w/\overline{a}}-1)} \frac{1}{w^2}.
    \label{R+ infty ultra}
\end{eqnarray}

\begin{figure}
    \centering
    \includegraphics[width=0.49\linewidth]{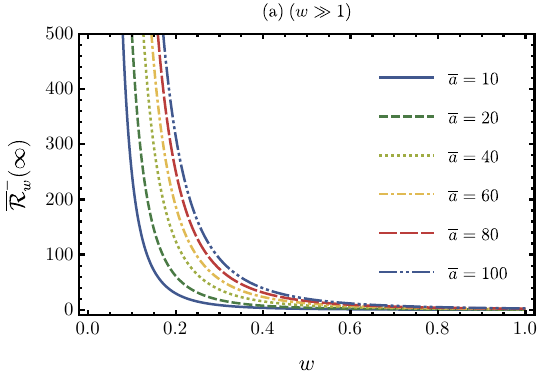}
    \includegraphics[width=0.49\linewidth]{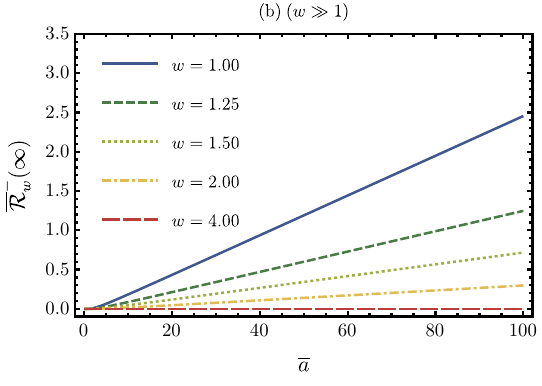}
    \caption{Excitation probability rate $\overline{\mathcal{R}}^{-}_{w}(\infty)$ for the regime where $w \gg 1$: \textbf{(a)} as a function of the parameter $w$ for different values of the parameter $\overline{a}$ and \textbf{(b)} as a function of the parameter $\overline{a}$ for different values of the parameter $w$.}
    \label{Fig0}
\end{figure}

In Fig~\ref{Fig0}, we plot the excitation probability rate $\overline{\mathcal{R}}^{-}_{w}(\infty)$ as a function of different parameters, in order to analyze the effect of the four-velocity component on the detector response behavior. Note that in both Fig.~\ref{Fig0}(a) and Fig.~\ref{Fig0}(b), as the parameter $w$ increases, the detector excitation rate decreases rapidly until it becomes zero. It is important to emphasize that for this regime, the detector response is suppressed even when considering small values for $w$ [see Fig.~\ref{Fig0}(a)]. Furthermore, see in Fig.~\ref{Fig0}(b) that for $w \geq 4.00$ the detector does not respond, and thus it is clear that using values of the type $w \gg 1$ the detector will not respond.

In this way, it is clarified through Eqs.~(\ref{R- infty ultra}) and (\ref{R+ infty ultra}), that the Unruh effect is suppressed when the four-component of the velocity reaches the ultra-relativistic regime ($w \gg 1$). It is important to emphasize that since the Unruh effect is suppressed in the ultra-relativistic regime, then the UDW detector will not respond, and consequently, we will not have the influence of acceleration radiation on accelerated quantum systems.

\section{\label{sec:level3}Velocity effects on an accelerated single-qubit}

\subsection{The theoretical model}

In particular, our goal is to clarify the role played by the velocity $w$ in shaping the fundamental characteristics of two-level quantum systems. To this end, we adopt a model in which a detector interacts linearly with a scalar quantum field. In this context, the field, immersed in a Minkowski spacetime background, is initially prepared in the vacuum state $\vert 0_{\mathcal{M}}\rangle$, while the detector is initialized in a general qubit state, namely,
\begin{eqnarray}
|\psi_{\mathrm{D}}\rangle = \alpha|g\rangle + \beta|e\rangle,
\label{qubitstate}
\end{eqnarray}
where $\alpha$ and $\beta$ are complex amplitudes given by $\alpha = \e^{i\frac{\varphi}{2}}\cos{\frac{\theta}{2}}$, $\beta = \e^{-i\frac{\varphi}{2}}\sin{\frac{\theta}{2}}$, and they satisfy the relationship $\vert\alpha\vert^2 + \vert\beta\vert^2=1$. See that here, $\theta \in [0,\pi]$ and $\varphi \in [0,2\pi]$ are the polar and azimuthal angles of the Bloch sphere \cite{jazaeri2019review, kasirajan2021quantum}. In this way, the density matrix $\hat{\rho}^{\mathrm{in}}_D = |\psi_D\rangle \langle\psi_D|$, using the Eq.~(\ref{qubitstate}), read as
\begin{eqnarray}
    \hat{\rho}^{\mathrm{in}}_D = |\alpha|^2 |g\rangle\langle g| + \alpha\beta^*|g\rangle\langle e| + \alpha^*\beta|e\rangle\langle g| + |\beta|^2|e\rangle\langle e|.
    \label{rhoinD}
\end{eqnarray}

We consider a UDW detector initially prepared in a qubit state [Eq.~(\ref{qubitstate})]. The detector moves along a trajectory within a two-dimensional plane, and couples linearly to a massless quantum field $\phi$ during a finite interaction interval $T$. After the interaction, one can measure the internal states $\vert g\rangle$ and $\vert e\rangle$, which provides a means to probe the modifications induced by the coupling. Formally, the initial state of the total system is given by $\hat{\rho}_{\mathrm{in}} = \hat{\rho}^{\mathrm{in}}_{D} \otimes \hat{\rho}_{\phi}$, with the field initialized in the Minkowski vacuum, i.e., $\hat{\rho}_{\phi} = |0_{\mathcal{M}}\rangle \langle 0_{\mathcal{M}}|$.

The density matrix after the interaction is governed by the Hamiltonian associated with the linear interaction [Eq.~(\ref{Hint})], and can be expressed as
\begin{eqnarray}
    \hat{\rho}^{\mathrm{out}} &=& \mathcal{\hat{U}}^{(0)} \hat{\rho}_{\mathrm{in}} \mathcal{\hat{U}}^{(0)^\dagger} + \mathcal{\hat{U}}^{(1)} \hat{\rho}_{\mathrm{in}} + \hat{\rho}_{\mathrm{in}}\mathcal{\hat{U}}^{(1)^\dagger} + \mathcal{\hat{U}}^{(1)} \hat{\rho}_{\mathrm{in}} \mathcal{\hat{U}}^{(1)^\dagger} + \mathcal{\hat{U}}^{(2)} \hat{\rho}_{\mathrm{in}} + \hat{\rho}_{\mathrm{in}} \mathcal{\hat{U}}^{(2)^\dagger} + \mathcal{O}(\lambda^3).\nonumber\\
\end{eqnarray}
In this manner, $\hat{\mathcal{U}}$ represents the time-evolution operator in its perturbative formulation, and can be explicitly expressed as
\begin{eqnarray}
    \mathcal{\hat{U}} = \mathcal{\hat{U}}^{(0)} + \mathcal{\hat{U}}^{(1)} + \mathcal{\hat{U}}^{(2)} + \mathcal{O}(\lambda^3),
    \label{U}
\end{eqnarray}
with the following perturbative terms
\begin{eqnarray}
    \mathcal{\hat{U}}^{(0)} &=& \mathbb{I},\\
    \mathcal{\hat{U}}^{(1)} &=& -i\lambda\int_{-\infty}^{\infty}d\tau\chi(\tau)\mu(\tau)\phi[x(\tau)],\\
    \mathcal{\hat{U}}^{(2)} &=& -\lambda^2\int^{+\infty}_{-\infty} d\tau \int^{+\tau}_{-\infty} d\tau' \chi(\tau)\chi(\tau')  \mu(\tau)\mu(\tau') \phi[x(\tau)]\phi[x(\tau')],
\end{eqnarray}
where $\mathbb{I}$ denotes the identity operator, and $\mu(\tau) = [\hat{\sigma}_{+}\e^{i\Omega\tau} + \hat{\sigma}_{-}\e^{-i\Omega\tau}]$, which promote transitions between the ground and excited states of the detector. Besides, the operators $\hat{\sigma}_{+} = |e\rangle\langle g|$ and $\hat{\sigma}_{-} = |g\rangle\langle e|$ are defined as the creation and annihilation operators, respectively.

We now turn our attention to the analysis of the final state of the UDW detector. In order to perform this investigation, it is necessary to trace over the degrees of freedom associated with the field configuration. This procedure yields $\hat{\rho}^{\mathrm{out}}_{D} = \mathrm{Tr}_{\vert0_{\mathcal{M}}\rangle}[\hat{\rho }^{\mathrm{out}}]$. The reduced density matrix for the detector results in the following expression:
\begin{eqnarray}
    \hat{\rho}^{\mathrm{out}}_{D} = \hat{\rho}^{\mathrm{in}}_{D} + \mathrm{Tr}_{\vert0_{\mathcal{M}}\rangle} \left(\mathcal{\hat{U}}^{(1)} \hat{\rho}_{\mathrm{in}} \mathcal{\hat{U}}^{(1)\dagger}\right) + \mathrm{Tr}_{\vert0_{\mathcal{M}}\rangle} \left(\mathcal{\hat{U}}^{(2)}\hat{\rho}_{\mathrm{in}}\right) + \mathrm{Tr}_{\vert0_{\mathcal{M}}\rangle} \left(\hat{\rho}_{\mathrm{in}}\mathcal{\hat{U}}^{(2)\dagger}\right),
    \label{rhooutD}
\end{eqnarray}
and defining the integrals as follows
\begin{eqnarray}
    \mathcal{C}^{\pm} &=& \int^{+\infty}_{-\infty} \d\tau \int^{+\infty}_{-\infty} \d\tau' \chi(\tau)\chi(\tau') \e^{\pm i\Omega(\tau+\tau')} \mathcal{W}(\tau, \tau'),
    \label{C+-}
\end{eqnarray}
\begin{eqnarray}
    \mathcal{G}^{\pm} &=& \int^{+\infty}_{-\infty} \d\tau \int^{+\tau}_{-\infty} \d\tau' \chi(\tau)\chi(\tau')  \e^{\pm i\Omega(\tau-\tau')} \mathcal{W}(\tau, \tau'),
    \label{intG}
\end{eqnarray}
\begin{eqnarray}
    \mathcal{F}^{\pm} &=& \int_{-\infty}^{+\infty} \d\tau \int_{-\infty}^{+\infty} \d\tau' \chi(\tau)\chi(\tau') \e^{\pm i\Omega(\tau - \tau')} \mathcal{W}(\tau, \tau'),
    \label{Flw}
\end{eqnarray}
after performing extensive and meticulous calculations, and omitting from now on the superscript ``$\mathrm{out}$'' to simplify notation, finally, we obtain that
\begin{eqnarray}
    \hat{\rho}_{D} &=& \left[ \cos^2{\frac{\theta}{2}} + \lambda^{2}\left( \mathcal{F}^{+}\sin^4{\frac{\theta}{2}} - \mathcal{F}^{-}\cos^4{\frac{\theta}{2}} \right) \right] |g\rangle \langle g| \nonumber\\
    &+& \left[ \sin^2{\frac{\theta}{2}} + \lambda^{2}\left( \mathcal{F}^{-}\cos^4{\frac{\theta}{2}} - \mathcal{F}^{+}\sin^4{\frac{\theta}{2}} \right) \right] |e\rangle \langle e| \nonumber\\
    &+& \left\{ \frac{1}{2}\e^{-i\varphi}\sin{\theta} + \frac{\lambda^{2}}{2} \left[\e^{i\varphi}\mathcal{C}^{+} - \e^{-i\varphi} \left(\mathcal{G}^{+} + \mathcal{G}^{-*}\right) \right]\sin{\theta} \right\} |e\rangle \langle g| \nonumber\\
    &+& \left\{ \frac{1}{2}\e^{+i\varphi}\sin{\theta} + \frac{\lambda^{2}}{2}  \left[\e^{-i\varphi}\mathcal{C}^{-} - \e^{i\varphi} \left(\mathcal{G}^{-} + \mathcal{G}^{+*}\right) \right] \sin{\theta} \right\} |g\rangle \langle e|.
    \label{final rho}
\end{eqnarray}
It should be emphasized that the reduced density matrix given in Eq.~(\ref{final rho}) must possess unit trace. In order to satisfy this requirement, the terms proportional to $\lambda^2$ cancel as a direct consequence of the normalization condition.
This is a consequential relation of the trace property of the density matrix.

\subsection{Quantum coherence}

Quantum coherence is a fundamental property of quantum systems, manifested through the existence of superposition states that enable interference between distinct eigenstates \cite{Leggett1980}. More precisely, it is characterized by the preservation of relative phases among the components of a superposed quantum state, thereby facilitating essential quantum phenomena such as interference and entanglement \cite{streltsov2015measuring}. 

Within this context, quantum optical methods represent a crucial set of techniques for the manipulation and control of coherence \cite{Glauber1963coherent, sudarshan1963}. In the particular case of a two-level quantum system, the coherence between the states $\vert g\rangle$ and $\vert e\rangle$ can be quantified by means of the $l^1$ norm quantum coherence, defined as the sum of the absolute values of the off-diagonal elements of the system’s density matrix \cite{Baumgratz2014QuantifyingCoherence}, namely:
\begin{align}
    \mathcal{Q}^{l^1}(\hat{\rho}_{D}) = \sum_{i \neq j} \mid \hat{\rho}_{D}^{ij}\mid,
    \label{DefCoherence}
\end{align}
and for the two-level system described by the density matrix given by Eq.~(\ref{final rho}), we obtain
\begin{align}
    \mathcal{Q}^{l^1} = \vert \sin{\theta}\vert \left\{ 1 - \lambda^2 \left[ \mathcal{F}^{+}\sin^2{\frac{\theta}{2}} + \mathcal{F}^{-}\cos^2{\frac{\theta}{2}} - \mathcal{C}^{-} \cos{(2\varphi)}\right]\right\} + \mathcal{O}(\lambda^{4}).
\end{align}
Using the relation $\mathcal{F}^{\pm} = \sigma\overline{\mathcal{R}}^{\pm}$, and considering a long interaction time\footnote{This regime is necessary to ensure that the detector has enough time to interact with the quantum field.}, namely $\sigma \gg 1$, we have
\begin{eqnarray}
    \mathcal{Q}^{l^1}_{w} = \vert \sin{\theta}\vert \left\{ 1 - \left[\frac{1}{2\pi} - w^2 F(\overline{a})(\e^{2\pi/\overline{a}} - 1)\right]\frac{\sigma\lambda^{2}}{\e^{2\pi/\overline{a}} - 1} \left( \cos^{2}{\frac{\theta}{2}} + \e^{2\pi/\overline{a}}\sin^{2}{\frac{\theta}{2}}\right)\right\},
\label{coherenceQubit_w}
\end{eqnarray}
where it is easy to see that by taking $w = 0$ we recover coherence for an accelerated single-qubit in one dimension, which reads as
\begin{eqnarray}
    \mathcal{Q}^{l^1}_{0} = \vert \sin{\theta}\vert \left\{ 1 - \frac{1}{2\pi}\frac{\sigma\lambda^{2}}{\e^{2\pi/\overline{a}} - 1} \left( \cos^{2}{\frac{\theta}{2}} + \e^{2\pi/\overline{a}}\sin^{2}{\frac{\theta}{2}}\right)\right\},
\label{coherenceQubit}
\end{eqnarray}
for more details on the expression of Eq.~(\ref{coherenceQubit}) see refs.~\cite{barros2024detecting, pedro2025mitigating} taking the corresponding limits for this.

\subsection{Numerical results}

\begin{figure}
    \centering
    \includegraphics[width=0.9\linewidth]{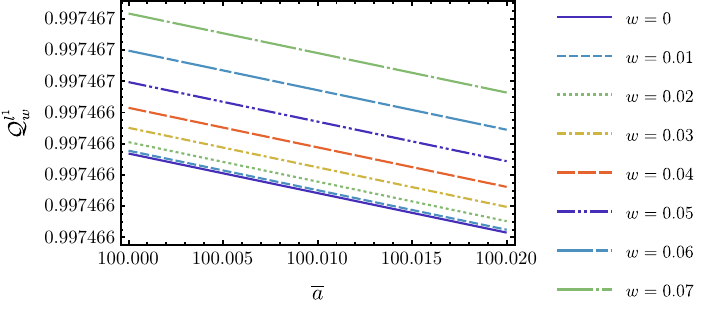}
    \caption{The $\mathcal{Q}^{l^1}_{w}$ as a function of the acceleration parameter $\overline{a}$ for different values of the four-velocity component $w$. We kept the following parameters constant: $\theta = \pi/2$, $\sigma = 10$, and $\lambda = 0.01$.}
    \label{Fig1}
\end{figure}

In this section, we discuss the numerical results obtained through the expression of the quantum coherence given by Eq.~(\ref{coherenceQubit_w}) of a accelerated single-qubit along a trajectory in a two-dimensional plane (in the non-relativistic regime, $w \ll 1$). Along this path, note that in Fig.~\ref{Fig1} we plot the $\mathcal{Q}^{l^1}_{w}$ as a function of the acceleration parameter $\overline{a}$ for different values of the four-velocity component $w$. As we increase the acceleration parameter $\overline{a}$, quantum coherence decays, indicating a coherence degradation  due to the Unruh effect. Furthermore, for increasing values of the parameter $w$, a slight increase in quantum coherence is observed, indicating that the effects of non-relativistic velocity slightly suppress the Unruh effect.

\begin{figure}
    \centering
    \includegraphics[width=0.9\linewidth]{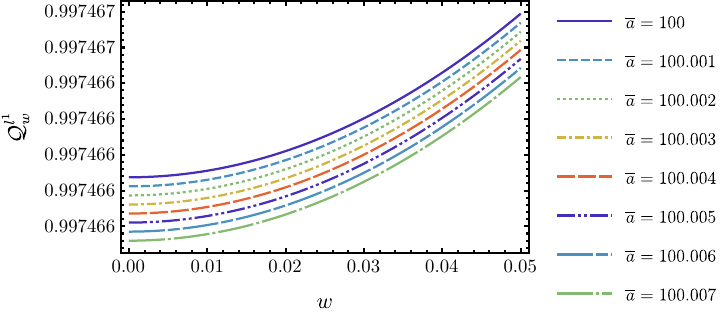}
    \caption{The $\mathcal{Q}^{l^1}_{w}$ as a function of the $w$ for different values of the acceleration parameter $\overline{a}$. We kept the following parameters constant: $\theta = \pi/2$, $\sigma = 10$, and $\lambda = 0.01$.}
    \label{Fig2}
\end{figure}

Similarly, in Fig.~\ref{Fig2} we plot the $\mathcal{Q}^{l^1}_{w}$ as a function of the four-velocity component $w$ for different values of the acceleration parameter $\overline{a}$. Note that as we increase the parameter $w$ the coherence also increases, furthermore, this implies that the non-relativistic motion of the detector in the $w$ direction mitigates the coherence degradation caused by the acceleration radiation (although these effects are very small, on the order of $10^{-6}$). On the other hand, when we increase the parameter $\overline{a}$ the quantum coherence decreases, this implies that even in the scenario of the detector moving in a two-dimensional plane we still have the degradation of coherence in the regime where $w$ is much smaller than the speed of light.

\begin{figure}
    \centering
    \includegraphics[width=0.9\linewidth]{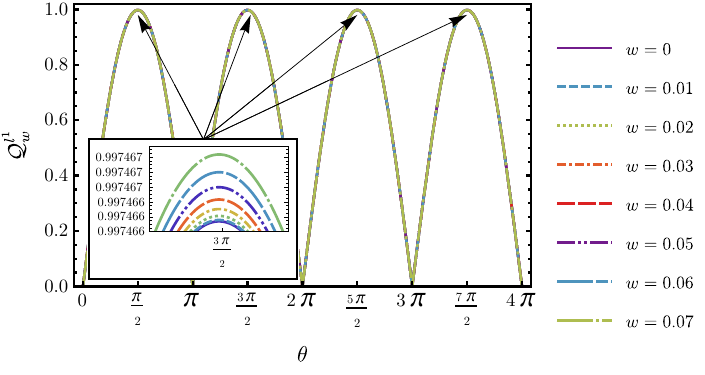}
    \caption{The $\mathcal{Q}^{l^1}_{w}$ as a function of the polar angle $\theta$ for different values of the four-velocity component $w$. We kept the following parameters constant: $\overline{a} = 100$, $\sigma = 10$, and $\lambda = 0.01$.}
    \label{Fig3}
\end{figure}

Additionally, in Fig~\ref{Fig3} we plot the $\mathcal{Q}^{l^1}_{w}$ as a function of the polar angle $\theta$ for different values of the four-velocity component $w$. The coherence amplitude attains its maximum value when $\theta$ is a half-integer multiple of $\pi$, which signifies the occurrence of maximal superposition between mixed states. Conversely, the coherence amplitude vanishes when $\theta$ is an integer multiple of $\pi$, corresponding to the poles of the Bloch sphere, where the qubit occupies a well-defined state devoid of superposition. Still in this plot, see that when we increase the values of the parameter $w$ the coherence amplitude also increases, again implying that the two-dimensional motion of this detector slightly mitigates coherence degradation.

\section{\label{sec:level4}Velocity effects on the duality relation}

Wave–particle duality constitutes a foundational principle of quantum theory and is quantitatively expressed through complementarity relations that connect interference visibility with path distinguishability \cite{Englert1996, wootters1979complementarity, greenberger1988simultaneous, bohr1928quantum, busch1985indeterminacy, Englert1999, Zeilinger1999Experiment}. In interferometric settings, the visibility provides a measure of the coherence between alternative quantum paths, whereas the distinguishability quantifies the degree of which-path information that may, in principle, be obtained by correlating the interfering system with an external degree of freedom \cite{Jaeger1995, Englert1996, miniatura2007path, kolavr2007path, bera2015duality}. These quantities satisfy well-established duality relations, which furnish an operational framework for analyzing the manner in which quantum coherence is diminished due to the transfer of information to the environment.

Within the framework of relativistic quantum systems, and particularly in the case of UDW detectors, this approach provides a natural generalization of the single-detector analysis developed in the preceding sections. Although the response of an individual detector already contains nontrivial information concerning motion-induced effects and field correlations \cite{Unruh1981Experimental, fulling1973, Davies_1975, Letaw1981stationary, abdolrahimi2014velocity}, the incorporation of the detector’s internal degree of freedom into a quantum interferometric circuit enables a direct investigation of how relativistic motion influences the interplay between coherence and information extraction \cite{Pedro2024robustness, barros2025information, pedro2025mitigating, Costa2020interferometry, GoodingUnruh2020interferometric, Lopes2021interference}. From this viewpoint, the interferometric formulation should not be regarded as an independent problem, but rather as an operationally transparent framework in which velocity-dependent modifications of the detector--field interaction are manifested through quantitative variations in interference visibility and path distinguishability, thereby naturally embedding the analysis of duality relations into the present study.

\subsection{Quantum interferometric circuit}

One indirect strategy for probing the properties of a physical system consists in the implementation of a quantum scattering circuit, which exploits quantum interferometry~\cite{Ramsey1950} to extract information about a specific subsystem by analyzing the measurable properties of an auxiliary probe. Quantum scattering circuits have found a wide range of applications, including tests of the Leggett--Garg inequality~\cite{souza2011scattering}, the measurement of correlation functions in simulations of the Fano--Anderson model~\cite{Negrevergne2005Liquid}, the determination of discrete Wigner functions~\cite{Leonhardt1995, miquel2002interpretation}, and the experimental reconstruction of work distributions~\cite{Batalhao2014}, among others. Such a circuit is characterized by a controlled interaction between a single qubit, which acts as a probe, and the system under investigation.

\begin{figure}
    \centering
    \includegraphics[width=0.97\linewidth]{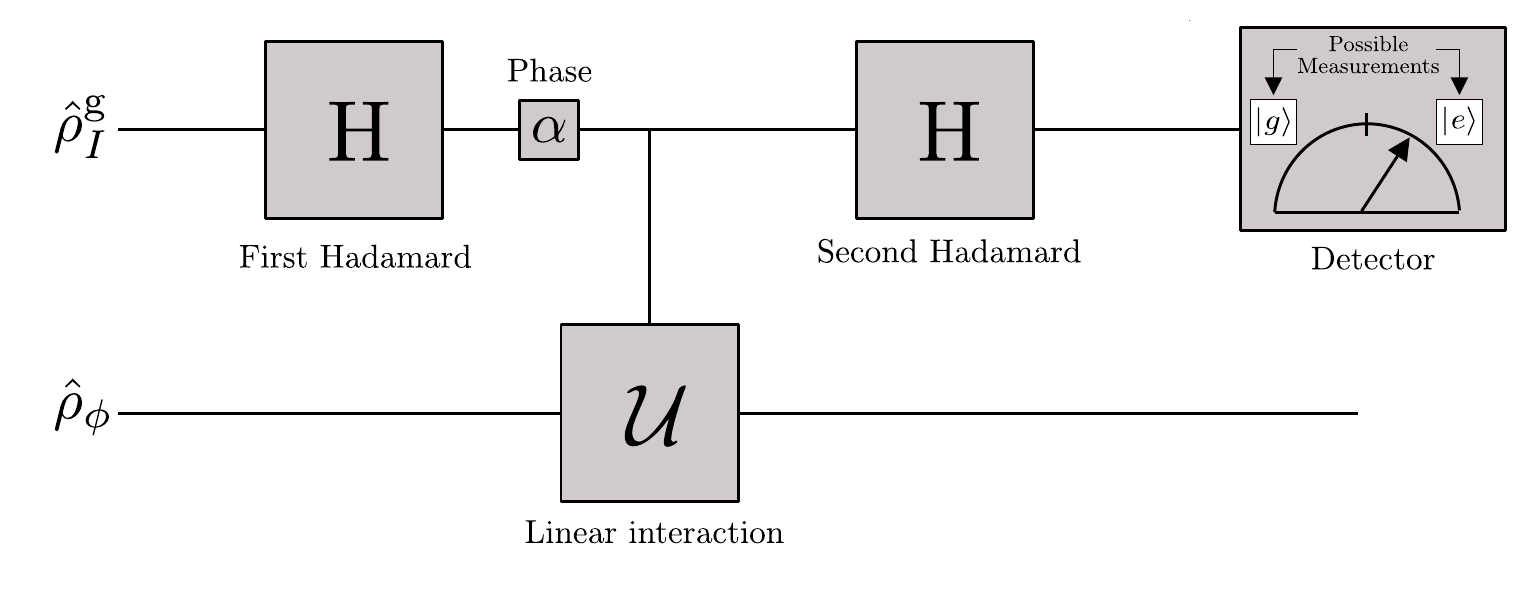}
    \caption{Schematic representation of the quantum interferometric circuit.}
    \label{Fig setup1}
\end{figure}

In this section, we investigate the quantum scattering circuit (see a schematic representation in Fig. ~\ref{Fig setup1}). The circuit is structured as follows: initially, a single qubit is prepared in a well-defined initial state. Subsequently, the application of the first Hadamard gate transforms the qubit into a coherent superposition. Thereafter, a phase accumulation takes place through the action of the phase-shift gate $\hat{\alpha}$. Following this step, the unitary operator $\hat{\mathcal{U}}$ [Eq.~(\ref{U})] induces a linear interaction, in a controlled manner, between the detector and the quantum field. Finally, the qubit is subjected to a second Hadamard operation, after which projective measurements on the system are performed.

In this configuration, the qubit utilized within the quantum interferometric circuit is initially described by the density operator $\hat{\rho}^{\mathrm{g}}_{D,I} = |g\rangle \langle g|$. Applying the first Hadamard gate to the initial state causes a superposition in it. During the evolution between the two Hadamard gates, the probability amplitudes associated with the qubit’s internal states undergo phase accumulation. In particular, the acquired phase is introduced by the phase-shift gate $\hat{\alpha}$ when the qubit is prepared in the excited state $|e\rangle$, i.e. $\hat{\alpha} = \frac{1}{\sqrt{2}}\,\text{diag}(1, \e^{i\alpha})$, and we have
\begin{eqnarray}
    \hat{\rho}^{\mathrm{in}}_{D,I} = \frac{1}{2}\left( |g\rangle \langle g| + \e^{-i\alpha}|g\rangle \langle e| + \e^{i\alpha}|e\rangle \langle g| + |e\rangle \langle e|\right).
\end{eqnarray}
Consequently, the initial state of the composite system can be expressed as $\hat{\rho}_{\mathrm{in},I} = \hat{\rho}^{\mathrm{in}}_I \otimes \hat{\rho}_\phi$. Therefore, the density operator describing the system after the interaction is given by
\begin{eqnarray}
    \hat{\rho}^{\mathrm{out}}_{I} &=& \mathcal{\hat{U}}^{(0)} \hat{\rho}_{\mathrm{in},I} \mathcal{\hat{U}}^{(0)^\dagger} + \mathcal{\hat{U}}^{(1)} \hat{\rho}_{\mathrm{in},I} + \hat{\rho}_{\mathrm{in},I}\mathcal{\hat{U}}^{(1)^\dagger} + \mathcal{\hat{U}}^{(1)} \hat{\rho}_{\mathrm{in},I} \mathcal{\hat{U}}^{(1)^\dagger} + \mathcal{\hat{U}}^{(2)} \hat{\rho}_{\mathrm{in},I} + \hat{\rho}_{\mathrm{in},I} \mathcal{\hat{U}}^{(2)^\dagger} + \mathcal{O}(\lambda^3).\nonumber\\
\end{eqnarray}
In this way, applying the partial trace over the degrees of freedom of the field, and then actuating the second Hadamard gate, we finally obtain (omitting the superscript ``out'')
\begin{eqnarray}
    \hat{\rho}_{I} &=& \Big[ \cos^2{\frac{\alpha}{2}} - \frac{\lambda^2}{4} \Big( \mathcal{F}^{-} + \mathcal{F}^{+} - 2\mathcal{C}^{-}  \Big)\cos{\alpha} \Big] |g\rangle \langle g| + \nonumber\\
    &+& \Big[ \sin^2{\frac{\alpha}{2}} + \frac{\lambda^2}{4} \Big( \mathcal{F}^{-} + \mathcal{F}^{+} - 2\mathcal{C}^{-}  \Big) \cos{\alpha} \Big] |e\rangle \langle e| +\nonumber\\
    &+&\Big\{ -\frac{i}{2}\sin{\alpha} + \frac{\lambda^2}{4} \Big[ \mathcal{F}^{-} - \mathcal{F}^{+} + 2i\sin{\alpha \left( 2\mathbf{Re}(\mathcal{G}^{-}) - \mathcal{C}^{-}\right)}\Big] \Big\} |e\rangle\langle g| + \nonumber\\
    &+& \Big\{ +\frac{i}{2}\sin{\alpha} + \frac{\lambda^2}{4} \Big[ \mathcal{F}^{-} - \mathcal{F}^{+} - 2i\sin{\alpha \left( 2\mathbf{Re}(\mathcal{G}^{-}) - \mathcal{C}^{-}\right)}\Big] \Big\} |g\rangle\langle e|.
    \label{rhoI final}
\end{eqnarray}
Note that the unitarity condition is satisfied. It is easy to see that the diagonal terms proportional to $\lambda^2$ vanish when we compute the trace of this matrix.

\subsection{Visibility and quantum coherence}

At this stage, we are in a position to extract information concerning the interference pattern produced by the quantum interferometric circuit. The interferometric visibility constitutes a quantitative metric for the contrast of the interference resulting from quantum superposition. Visibility is formally defined as the ratio between the amplitude of the interference fringes and the sum of the corresponding individual intensities. Mathematically, this is expressed as:
\begin{eqnarray}
    \mathcal{V}_{I} = \frac{P^{\mathrm{g,\,max}}_{I} - P^{\mathrm{g,\,min}}_{I}}{P^{\mathrm{g,\,max}}_{I} + P^{\mathrm{g,\,min}}_{I}}.
\end{eqnarray}
where $P^{\mathrm{g}}_{I} = \langle g|\hat{\rho}_{I}|g\rangle$ is the probability of finding the ground state. In this context, the probability $P^{\mathrm{g}}_{I}$ attains its maximum (minimum) value when $\alpha = 0$ ($\alpha = \pi$). Accordingly, by taking into account the limit of long interaction times ($\sigma \gg 1$), we obtain:
\begin{eqnarray}
    \mathcal{V}_{I,w} &\approx& 1 - \left[1 - 2\pi\,w^2 F(\overline{a})(\e^{2\pi/\overline{a}} - 1)\right]\frac{\sigma \lambda^2}{4\pi} \coth{\left(\frac{\pi}{\overline{a}}\right)},
    \label{visibility_w}
\end{eqnarray}
this result represents the interferometric visibility with the velocity effects, in this way, taking $w = 0$, we reproduce the result for the case of a trajectory with acceleration in one dimension, and therefore we have
\begin{eqnarray}
    \mathcal{V}_{I,0} &\approx& 1 - \frac{\sigma \lambda^2}{4\pi} \coth{\left(\frac{\pi}{\overline{a}}\right)}.
    \label{visibility}
\end{eqnarray}

In this way, we can obtain the $l^1$ norm quantum coherence for the case of the quantum interferometric circuit, which can be easily verified that it is read as $\mathcal{Q}^{l^1}_{I,w} \approx \mathcal{V}_{I,w}\,\sin{\alpha}$, and mathematically we have
\begin{eqnarray}
    \mathcal{Q}^{l^1}_{I,w} &\approx& \sin{\alpha} \left\{1 - \left[1 - 2\pi\,w^2 F(\overline{a})(\e^{2\pi/\overline{a}} - 1)\right]\frac{\sigma \lambda^2}{4\pi} \coth{\left(\frac{\pi}{\overline{a}}\right)}\right\},
    \label{coherence_w}
\end{eqnarray}
where it is also valid to take the limit $w = 0$ to obtain coherence for the case of an interferometer without velocity effects.

\subsection{Which-path distinguishability circuit}

\begin{figure}
    \centering
    \includegraphics[width=0.99\linewidth]{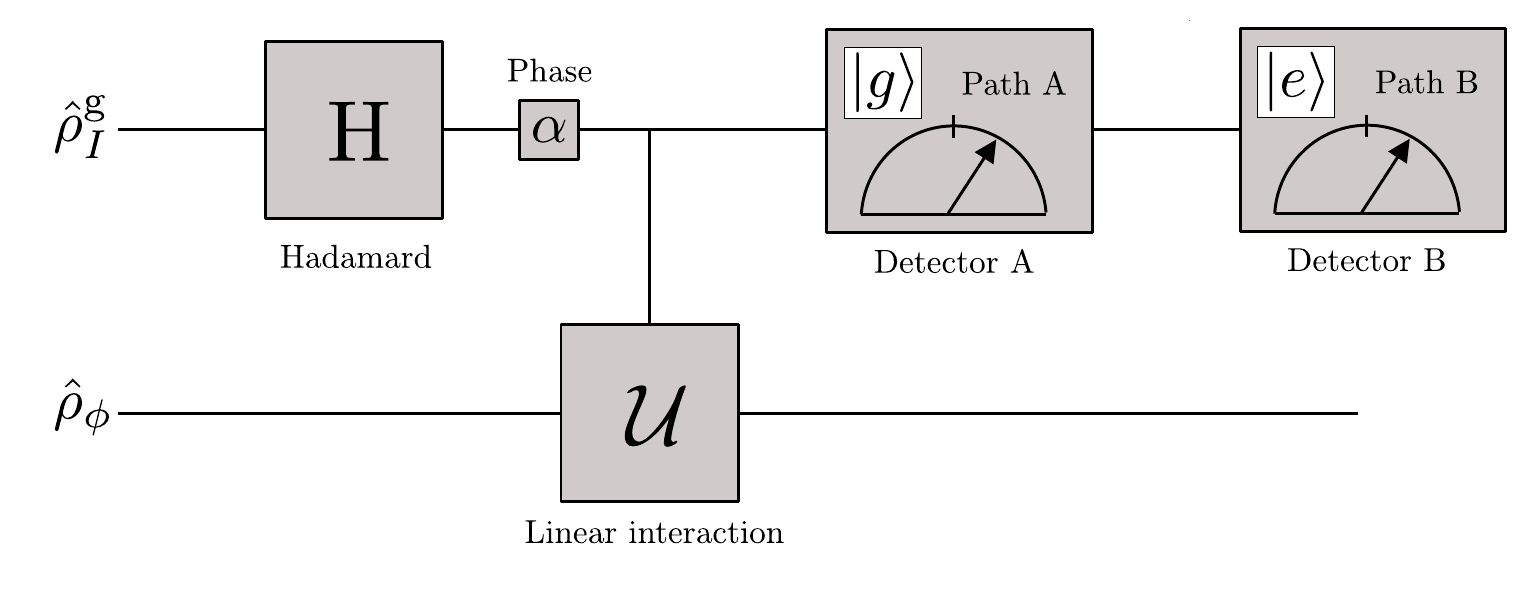}
    \caption{Schematic representation of the which-path distinguishability circuit.}
    \label{Fig setup2}
\end{figure}

Now, we consider a modified configuration aimed at extracting which-path information. Specifically, the second Hadamard gate is removed from the quantum interferometric circuit, and two detectors are incorporated (as shown in Fig. ~\ref{Fig setup2}). The introduction of the path detectors allows the interaction between the qubit and a massless scalar field to reveal the trajectory followed by the qubit. Specifically, the internal state of the qubit is initialized in the state $|g\rangle$ for qubits registered along path A by detector A. Conversely, qubits that are not registered by detector A are inferred to have propagated along path B. In this manner, the which-path information is effectively encoded in the internal states of the qubit.

A central quantity in this system is the which-path distinguishability, which characterizes the particle-like behavior of the qubit. This quantity is formally quantified by the following expression:
\begin{equation}
 \mathcal{D} = \frac{|w_{A} - w_{B}|}{w_{A} + w_{B}},
 \label{Dis1}
\end{equation}
where $w_{A}$ denotes the probability of observing the qubit at detector A (associated with path A), and $w_{B}$ corresponds to the probability of its detection at detector B (associated with path B), namely we have
\begin{eqnarray}
 w_{A} &=& \frac{1}{2} + \frac{\lambda^2}{2} \left(\mathcal{F}^- - \mathcal{F}^+\right), \\
 w_{B} &=& \frac{1}{2} + \frac{\lambda^2}{2} \left(\mathcal{F}^+ - \mathcal{F}^-\right).
\end{eqnarray}
By using $\mathcal{F}^\pm = \sigma \overline{\mathcal{R}}^\pm$ and Eq. \eqref{Dis1}, using $\sigma \ll 1$, the which-path distinguishability is
\begin{equation}
 \mathcal{D}_{w} = \left[\frac{1}{2\pi} - w^2 F(\overline{a})(\e^{2\pi/\overline{a}} - 1)\right]\sigma \lambda^2 + \mathcal{O}(\lambda^{4}),
 \label{DisFinal_w}
\end{equation}
and taking $w = 0$ we then have $\mathcal{D}_{0} = \frac{\sigma \lambda^2}{2\pi} + \mathcal{O}(\lambda^{4})$ which is the expression for the which-path distinguishability without velocity effects. It is important to note that the velocity effects introduces an acceleration dependence on path information.

\subsection{Complementarity relation}

The wave–particle duality \cite{wootters1979complementarity, greenberger1988simultaneous} encapsulates the intrinsic trade-off between these two physical attributes. In particular, any attempt to obtain which-path information (i.e., an increase in the distinguishability $\mathcal{D}_{w}$) necessarily leads to a reduction in the coherence of the interference pattern (i.e., a decrease in the visibility $\mathcal{V}_{I,w}$), and conversely \cite{bohr1928quantum, busch1985indeterminacy}. This fundamental trade-off is rigorously formalized by the inequality originally derived in the seminal works of Englert and Zeilinger \cite{Englert1996, Englert1999, Zeilinger1999Experiment}, expressed as
\begin{equation} 
    C_{w} = \mathcal{V}^2_{I,w} + \mathcal{D}^2_{w} \leq 1.
\label{Complementarity1_w}
\end{equation}
Besides, this duality is typically demonstrated in interference experiments~\cite{menzel2012wave, wang2021molecular, yoon2021quantitative, chen2022experimental}. Substituting Eqs. (\ref{visibility_w}) and (\ref{DisFinal_w}) into (\ref{Complementarity1_w}), we concluded
\begin{eqnarray}
    C_{w} &\approx& 1 - \left[\frac{1}{2\pi} - w^2 F(\overline{a})(\e^{2\pi/\overline{a}} - 1)\right]\sigma \lambda^2 \coth{\left(\frac{\pi}{\overline{a}}\right)},
    \label{ComplementarityFinal_w}
\end{eqnarray}
and taking $w = 0$, we have
\begin{eqnarray}
    C_{w} &\approx& 1 - \frac{\sigma \lambda^2}{2\pi} \coth{\left(\frac{\pi}{\overline{a}}\right)},
    \label{ComplementarityFinal}
\end{eqnarray}
and we obtain the complementarity relation for the case without velocity effects.

\subsection{Numerical results}

\begin{figure}
    \centering
    \includegraphics[width=0.9\linewidth]{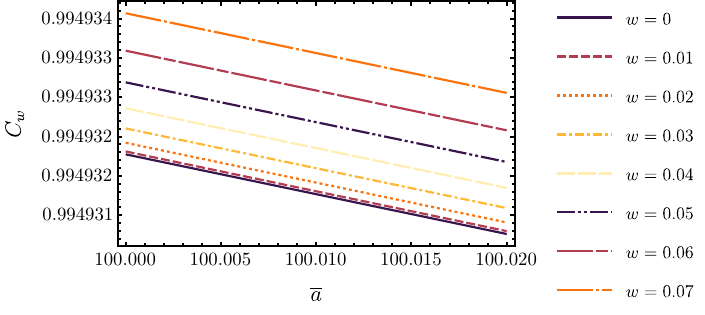}
    \caption{The behavior of $C_{w}$ is presented as a function of the acceleration parameter $\overline{a}$ for distinct values of the four-velocity component $w$. Throughout the analysis, the following parameters were kept fixed: $\sigma = 10$, and $\lambda = 0.01$.}
    \label{Fig4}
\end{figure}

Now, in this section we present the numerical results of the effects of non-relativistic velocity ($w \ll 1$) on the complementarity relation [Eq.~(\ref{ComplementarityFinal_w})] obtained through interferometric visibility [Eq.~(\ref{visibility_w})] and which-path distinguishability [Eq.~(\ref{DisFinal_w})]. Through this analysis, it is possible to understand the wave-particle duality behavior of the system. In Fig.~\ref{Fig4} we plot $C_{w}$ as a function of the $\overline{a}$ for distinct values of the $w$. Note that as we increase the acceleration the complementarity relation decays, this occurs due to the effects caused by Unruh radiation that degrade the wave-particle information of systems at high accelerations~\cite{Pedro2024robustness, pedro2025mitigating, barros2025information}. On the other hand, when we take increasing values of $w$ we see that the degradation of this information is mitigated slightly.

\begin{figure}
    \centering
    \includegraphics[width=0.9\linewidth]{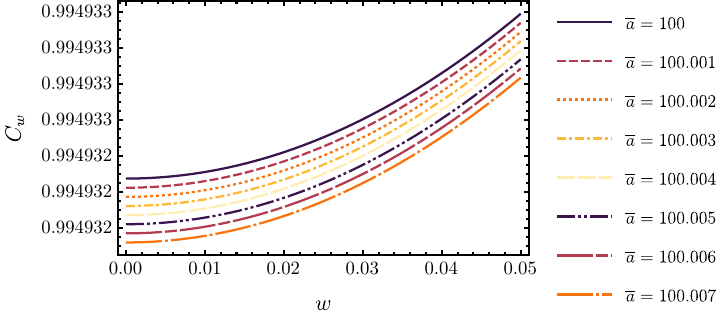}
    \caption{The $C_{w}$ as a function of $w$ is analyzed for different values of the acceleration parameter $\overline{a}$. The following parameters were held fixed throughout the analysis: $\sigma = 10$, and $\lambda = 0.01$.}
    \label{Fig5}
\end{figure}

In this way, in Fig.~\ref{Fig5} we plot the $C_{w}$ as a function of $w$ is analyzed for different values of the acceleration parameter $\overline{a}$. Note that as we increase the four-velocity component $w$, the complementarity ratio also increases, and once again, these results show that the constant, non-relativistic motion of the detector in the direction of $w$ mitigates the degradation of wave-particle information caused by detector acceleration. Note also that the effects of this mitigation are quite small, but they are important since they demonstrate a way to protect a system's information at high accelerations.

\section{\label{sec:level5}Model limitations}

In this section, we present the assumptions and limitations of our theoretical model studied in this paper. First, we employed perturbation theory to analyze the detector-field interaction. This method relies on the assumption of a small coupling constant ($\lambda \ll 1$), which guarantees the validity of the perturbative expansion, but simultaneously restricts the analysis to regimes of weak interactions. Moreover, we assumed that the interaction takes place over a finite yet sufficiently long time interval ($\sigma \gg 1$) when compared to the characteristic transition time of the two-level system ($\sim \Omega^{-1}$), this assumption ensures that the detector has enough time to interact with the quantum field. It is important to emphasize that, in different parameter regimes, deviations may arise that make it difficult to generalize our results to arbitrary time scales.

We considered the high-acceleration regime ($\overline{a} \gg 1$), in such a situation the detector acceleration far exceeds the transition frequency $\Omega$ and guarantees the presence of the Unruh effect. It is important to emphasize that the model neglects the back-reaction effects, and higher-order corrections and non-perturbative contributions are disregarded. Regarding the assumptions and limitations of velocity effects, in this paper we consider the asymptotic limits, namely: non-relativistic velocity ($w \ll 1$) and the ultra-relativistic velocity ($w \gg 1$). We saw previously that for the ultra-relativistic regime the detector under high accelerations does not respond because the Unruh effect is suppressed, and because of this it is not possible to see the signature of acceleration radiation in the quantum systems studied here. On the other hand, for the non-relativistic regime, we found significant effects of the four-velocity component $w$. We saw that these effects are quite small for all quantum systems studied here at high accelerations. This is certainly because the scale of the $w$ component is much smaller than the scalar of the acceleration parameter $\overline{a}$.

\section{\label{sec:level6}Summary and conclusion}

In this work, we focus on the investigation of a UDW detector following a world line described in a two-dimensional plane as shown by Eq.~(\ref{trajectory}), in order to know the influence of the four-velocity component $w = \d y/\d\tau$ (constant) on the degradation of the information of quantum systems at high accelerations. Furthermore, our study was based on perturbation theory and considered a finite interaction time that results in slight modifications of the thermal spectrum of the detector response.

Evidently, in the expressions for the non-relativistic velocity regime, the presence of terms proportional to $w^{2}$ was observed for all quantum systems studied, which serves as a witness to non-relativistic velocity effects. On the other hand, for the ultra-relativistic velocity regime we observe that the transition rates are proportional to $w^{-4}$ and this causes the Unruh effect to be suppressed, and consequently, we do not see any effects in the accelerated quantum systems. It should be emphasized that an UDW detector following such a trajectory, in the ultrarelativistic limit, exhibits the same effect as that of a system interacting with its environment through the presence of boundaries under certain circumstances \cite{yang2018quantum, Jin2015Electromagnetic, liu2016protecting, liu2016inhibiting}. It is known that the total suppression of the detector response depends on many conditions, therefore, our findings depend on the model and do not have a general character.

Furthermore, the numerical results show that for all studied quantum systems, the information degradation decreases as we increase the parameter $w$, implying that the four-velocity component under very sensitive conditions serves as a protective agent for the information of these systems. This interesting behaviors happen because of the composite effect of both velocity and acceleration \cite{abdolrahimi2014velocity,  Liu2021relativistic, Hao2023Quantum, mukherjee2024enhancement, xie2025weak}, and in this work we witness these effects for other systems. More specifically, our findings showed that both for an accelerated single-qubit and for interferometric circuits, velocity effects cause mitigation in the degradation of the quantumness of these quantum systems in the high acceleration regime.

Therefore, the non-relativistic transverse motion in the $y$-direction modifies the detector response spectrum, reducing the Unruh radiation responsible for the degradation of information associated with acceleration in the $x$-direction. This effect implies that transverse dynamics can act as a quantumness mitigation mechanism. However, the effects found are very small (on the order of $10^{-6}$) and should not be interpreted as robust protection of quantum information. The main relevance of the study lies in its conceptual character, revealing that specific trajectories can subtly alter the detector response and indicate little-explored theoretical pathways in RQI. We emphasize that we do not claim immediate experimental viability, but rather the value of the model as a theoretical exercise capable of expanding understanding of the effects of transverse dynamics in accelerated systems.

\begin{acknowledgments}
P.H.M.~Barros acknowledges the Brazilian funding agency CAPES for financial support through grant No.~88887.674765/2022-00 (Doctoral Fellowship – CAPES);  S.M.~Wu was supported by the National Natural Science Foundation of China (Grant Nos.~12205133) and the Special Fund for Basic Scientific Research of Provincial Universities in Liaoning under grant No.~LS2024Q002. C.A.S. Almeida is supported by grant number 309553/2021-0 (CNPq/PQ) and by Project UNI-00210-00230.01.00/23 (FUNCAP). The authors thank the Referee for helpful comments.
\end{acknowledgments}

\appendix

\section{Derivation of Wightman function expansions\label{App:A}}

In this appendix section we explicitly show the derivations of the asymptotic expansions of the Wightman function given by Eq.~(\ref{wightman function}), presenting the two expansions, namely: for very small $w$ and for very large $w$.

\subsection{Non-relativistic regime}

Considering that the component of the four-velocity $w$ is very small, that is, $w \ll 1$, then the non-relativistic regime is obtained. Therefore, for these conditions using the binomial expansion, we have that Eq.~(\ref{alpha}) can be written as $\alpha \approx a[1 - \frac{w^2}{2} + \mathcal{O}(w^4)]$, and substituting this expression into the Wightman function Eq.~(\ref{wightman function}), we obtain
\begin{eqnarray}
    \mathcal{W}_{w}^{+}(\Delta\tau) \approx -\frac{a^2(1-\frac{w^2}{2})^4}{16\pi^2} \Bigg\{ \sinh^2{\Bigg[ \frac{a}{2}\left( 1 - \frac{w^2}{2}\right) \left( \Delta\tau - 2i\epsilon \right)\Bigg]} - \frac{w^2(a\Delta\tau)^2(1-\frac{w^2}{2})^4}{4} \Bigg\}^{-1}.\nonumber\\
\end{eqnarray}
Using the expansion given by $\left( 1 - \frac{w^2}{2}\right)^4 \approx 1 - 2w^2 + \mathcal{O}(w^4)$, and disregarding the terms $\mathcal{O}(w^4)$, we have
\begin{eqnarray}
    \mathcal{W}_{w}^{+}(\Delta\tau) \approx -\frac{a^2(1-2w^2)}{16\pi^2} \Bigg\{ \sinh^2{\Bigg[ \frac{a}{2} \left( \Delta\tau - 2i\epsilon \right) - \frac{a\,w^2}{4}\left( \Delta\tau - 2i\epsilon \right)\Bigg]} - \frac{w^2(a\Delta\tau)^2}{4} \Bigg\}^{-1}.\nonumber\\
    \label{wightman function app}
\end{eqnarray}
Now, to proceed, we can make the following definitions:
\begin{eqnarray}
    A \equiv \frac{a}{2}\left( \Delta\tau - 2i\epsilon \right), \qquad \text{and} \qquad B \equiv \frac{a\,w^2}{4}\left( \Delta\tau - 2i\epsilon \right).
\end{eqnarray}
Thus, we can analyze the following hyperbolic trigonometric relationship,
\begin{eqnarray}
    \sinh{(A - B)} = \sinh{(A)} \cosh{(B)} - \cosh{(A)}\sinh{(B)},
\end{eqnarray}
where $B$ is very small, since it is proportional to $w^2$, thus the following approximations are valid:
\begin{eqnarray}
    \cosh{(B)} \approx 1 + \frac{B^2}{2}, \qquad \text{and} \qquad \sinh{(B)} \approx B.
\end{eqnarray}
Thus, we have
\begin{eqnarray}
    \sinh{(A - B)} &\approx& \sinh{(A)} \left( 1 + \frac{B^2}{2} \right) - B\cosh{(A)},\\
    &\approx& \sinh{(A)} + \frac{B^2}{2}\sinh{(A)}  - B\cosh{(A)},
\end{eqnarray}
where $B^2 \sim \mathcal{O}(w^4)$, and therefore, we obtain
\begin{eqnarray}
    \sinh{(A - B)} &\approx& \sinh{(A)} \left( 1 + \frac{B^2}{2} \right) - B\cosh{(A)},\\
    &\approx& \sinh{(A)} - B\cosh{(A)} + \mathcal{O}(w^4).
\end{eqnarray}
Now, we can square both sides of this expression, disregarding the terms $\mathcal{O}(w^4)$, and thus we obtain:
\begin{eqnarray}
    \sinh^2{(A - B)} \approx \sinh^2{(A)} - 2B\sinh{(A)}\cosh{(A)} + \mathcal{O}(w^4),
\end{eqnarray}
where using the hyperbolic trigonometric relation given by $\sinh{(2A)} = 2\sinh{(A)}\cosh{(A)}$, we have
\begin{eqnarray}
    \sinh^2{(A - B)} \approx \sinh^2{(A)} - B\sinh{(2A)} + \mathcal{O}(w^4).
    \label{sinh relation}
\end{eqnarray}

Now, substituting the relation given by Eq.~(\ref{sinh relation}) into Eq.~(\ref{wightman function app}), we have that the Wightman function is written as
\begin{eqnarray}
    \mathcal{W}_{w}^{+}(\Delta\tau) &\approx& -\frac{a^2(1-2w^2)}{16\pi^2} \Bigg\{ \sinh^2{\left(  \frac{a\Delta\tau}{2} - i\epsilon a\right)}  - w^2\left( \frac{\Delta\tau}{4} - i\epsilon a\right)\sinh{\left(  a\Delta\tau - 2i\epsilon a\right)} + \nonumber\\
    &-& \frac{w^2(a\Delta\tau)^2}{4} \Bigg\}^{-1} + \mathcal{O}(w^4). 
\end{eqnarray}
To proceed with the mathematical manipulations in a clearer and simpler way, we can make the following definitions:
\begin{eqnarray}
    D &\equiv& \sinh^2{\left(  \frac{a\Delta\tau}{2} - i\epsilon a\right)}, \label{def D}\\
    E &\equiv& w^2\left( \frac{\Delta\tau}{4} - i\epsilon a\right)\sinh{\left(  a\Delta\tau - 2i\epsilon a\right)}, \label{def E}\\
    F &\equiv& \frac{w^2(a\Delta\tau)^2}{4}, \label{def F}
\end{eqnarray}
and with these definitions, our Wightman function now depends on the following term, $(D - E - F)^{-1}$, on which we can perform the following manipulations:
\begin{eqnarray}
    \frac{1}{D - E - F} = \frac{1}{D - E - F}\left(\frac{D + E + F}{D + E + F}\right) = \frac{D + E + F}{D^2 - (E + F)^2}.
\end{eqnarray}
Note that the term $(B + C)^2$ is of the order of $\mathcal{O}(w^4)$ and therefore we can disregard it, and thus we can continue with the following algebraic manipulations:
\begin{eqnarray}
    \frac{1}{D - E - F} \approx \frac{D + E + F}{A^2} + \mathcal{O}(w^4) \approx \frac{1}{D} + \frac{E + F}{D^2} + \mathcal{O}(w^4),
\end{eqnarray}
and replacing the definitions given by Eqs.~(\ref{def D}), (\ref{def E}), and (\ref{def F}), the Wightman function can be read as
\begin{eqnarray}
    \mathcal{W}_{w}^{+}(\Delta\tau) &\approx& -\frac{a^2(1-2w^2)}{16\pi^2} \Bigg\{ \frac{1}{\sinh^2{\left(  \frac{a\Delta\tau}{2} - i\epsilon a\right)}}  + \Bigg[w^2\left( \frac{\Delta\tau}{4} - i\epsilon a\right)\sinh{\left(  a\Delta\tau - 2i\epsilon a\right)} + \nonumber\\
    &+& \frac{w^2(a\Delta\tau)^2}{4}\Bigg]\frac{1}{\sinh^4{\left(  \frac{a\Delta\tau}{2} - i\epsilon a\right)}} \Bigg\} + \mathcal{O}(w^4). 
\end{eqnarray}

Finally, we have the expansion of the Wightman function for $w \ll 1$, and it is written as
\begin{eqnarray}
    \mathcal{W}_{w}^{+}(\Delta\tau) &\approx& -\frac{a^2}{16\pi^2} \Bigg\{ \frac{(1-2w^2)}{\sinh^2{\left(  \frac{a\Delta\tau}{2} - i\epsilon a\right)}}  + \Bigg[\left( \frac{\Delta\tau}{4} - i\epsilon a\right)\sinh{\left(  a\Delta\tau - 2i\epsilon a\right)} + \nonumber\\
    &+& \frac{(a\Delta\tau)^2}{4}\Bigg]\frac{w^2}{\sinh^4{\left(  \frac{a\Delta\tau}{2} - i\epsilon a\right)}} \Bigg\} + \mathcal{O}(w^4),
    \label{Wightman function w<<1}
\end{eqnarray}
where in the last step we multiply the term $(1 - 2w^2)$ across all terms within parentheses and disregard the term $\mathcal{O}(w^4)$.

\subsection{Ultra-relativistic regime}

Now, in this section we derive in detail the expansion of the Wightman function for the ultra-relativistic regime, that is, when $w \gg 1$. Under these conditions, it is easy to see that for very large speeds the parameter $\alpha = \frac{a}{\sqrt{1 + w^2}}$ can now be written as $\alpha \approx \frac{a}{w}$, and applying this to Eq.~Eq.~(\ref{wightman function}), we obtain
\begin{eqnarray}
    \mathcal{W}_{w}^{+}(\Delta\tau) &\approx& -\frac{a^2}{16\pi^2}\frac{1}{w^4} \Bigg[ \sinh^2{\left(  \frac{a\Delta\tau}{2w} - \frac{i\epsilon a}{w^2}\right)}  - \frac{(a\Delta\tau)^2}{4w^2}\Bigg]^{-1},
\end{eqnarray}
and to proceed, we can make the following definitions:
\begin{eqnarray}
    G &\equiv& \sinh^2{\left(  \frac{a\Delta\tau}{2w} - \frac{i\epsilon a}{w^2}\right)}, \label{def G}\\
    H &\equiv& \frac{(a\Delta\tau)^2}{4w^2}. \label{def H}
\end{eqnarray}
Through these definitions, it is possible to note that the Wightman function now depends on the term given by
\begin{eqnarray}
    \frac{1}{w^4}\frac{1}{G - H} = \frac{1}{w^4}\frac{1}{G - H} \left( \frac{G+H}{G+H}\right) = \frac{G + H}{w^4G^2 - w^4H^2},
\end{eqnarray}
where $w^4H^2$ is very small, since it is proportional to $w^{-8}$ and this term can be disregarded. Thus, we obtain
\begin{eqnarray}
    \frac{1}{w^4}\frac{1}{G - H} \approx \frac{G + H}{w^4G^2} \approx \frac{1}{w^4G} + \frac{H}{w^4G^2},
\end{eqnarray}
Note that the term $\frac{H}{w^4 G^2}$ is proportional to $w^{-6}$, and we can disregard it as well, and therefore we have
\begin{eqnarray}
    \mathcal{W}_{w}^{+}(\Delta\tau) &\approx& -\frac{a^2}{16\pi^2}\frac{1}{w^4} \Bigg[ \sinh^2{\left(  \frac{a\Delta\tau}{2w} - \frac{i\epsilon a}{w^2}\right)}\Bigg]^{-1}  + \mathcal{O}(w^{-6}),
\end{eqnarray}
and this expression is the expansion of the Wightman function to the ultra-relativistic regime.

\section*{Conflicts of interest/competing interest}
The authors declared that there is no conflict of interest in this manuscript.

\section*{Data Availability Statement}
This article has no associated data or the data will not
be deposited.

\section*{Code Availability Statement}
This article has no associated code or the code will not
be deposited.

\bibliography{main}
\bibliographystyle{unsrt}

\end{document}